\documentclass[usenatbib]{emulateapj} 
\usepackage{apjfonts}
\usepackage{graphicx}

\newcommand{\be}{\begin{equation}}
\newcommand{\ee}{\end{equation}}
\newcommand{\ba}{\begin{eqnarray}}
\newcommand{\ea}{\end{eqnarray}}

\shorttitle{The Environmental Dependencies of Giants and Dwarfs}
\shortauthors{Haines et al.}
\begin{document}

\title{The Different Environmental Dependencies of Star-formation for Giant and Dwarf Galaxies}

\author{C. P. Haines,
 F. La Barbera, A. Mercurio, P. Merluzzi,
and G. Busarello}
\affil{INAF - Osservatorio Astronomico di Capodimonte,
via Moiariello 16, I-80131 Napoli, Italy; chris@na.astro.it}


\begin{abstract}
We examine the origins of the bimodality observed in the global properties
of galaxies around a stellar mass of \mbox{$3\!\times\!10^{10}{\rm M}_{\sun}$} by
comparing the environmental dependencies of star-formation for the
giant and dwarf galaxy populations.
The Sloan Digital Sky Survey DR4 spectroscopic dataset is used to produce a sample of galaxies in the vicinity of the supercluster
centered on the cluster A2199 at $z=0.03$ that is \mbox{$\ga\!9$0\%} complete to a magnitude limit of
\mbox{${\rm M}_{r}^{*}+3.3$}. From these we measure global trends with
environment for both giant \mbox{(M$_{r}\!<\!-20$ mag)}
and dwarf \mbox{($-19<{\rm M}_{r}\!<\!-17.8$ mag)} subsamples using
the luminosity-weighted mean stellar age and
H$\alpha$ emission as independent measures of star-formation history. 
The fraction of giant galaxies classed as old 
(\mbox{$t>\!7$\,Gyr}) or passive (\mbox{EW$[{\rm H}\alpha]\le4$\,\AA})
falls gradually from $\ga\!8$0\% in the cluster cores to
\mbox{$\sim\!4$0\%} in field regions
beyond \mbox{3--4\,$R_{vir}$}, as found in previous studies.
In contrast, we find that the dwarf galaxy population shows a sharp
transition at \mbox{$\sim\!1\,R_{\rm vir}$}, from being predominantly 
old/passive within the cluster, to outside where virtually all galaxies
are forming stars and old/passive galaxies are {\em only} found as satellites to more massive galaxies. 
These results imply fundamental differences in the 
evolution of giant and dwarf galaxies: 
whereas the star-formation histories of giant galaxies are determined
primarily by their merger history, star-formation in dwarf galaxies is
much more resilient to the effects of major mergers. Instead dwarf
galaxies become passive only once
 they become satellites within a more massive halo, by losing their halo
 gas reservoir to the host halo, or through other environment-related
 processes such as galaxy harassment and/or
 ram-pressure stripping.  

\end{abstract}


\section{Introduction}
\label{intro}

The star-formation histories, masses and structural properties of galaxies,
are strongly dependent on their environment: massive,
passively-evolving spheroids dominate cluster cores, whereas in field
regions galaxies are typically low-mass, star-forming and
disk-dominated \citep[e.g.][hereafter K04]{blanton05,k04}. The Sloan Digital Sky Survey (SDSS) has allowed these environmental dependences
to be studied in detail
\citep[e.g.][]{gomez,tanaka04}, showing that for
massive galaxies at least \mbox{($\la\!{\rm M}^{*}+1$ mag)}, star-formation is
most closely dependent on local density, independent of the
richness of the nearest cluster or group, and is still
systematically suppressed for regions as far as 3--4
virial radii  \mbox{(${\rm R}_{vir}$)} from the nearest cluster. Hence the evolution of massive galaxies is
not primarily driven by mechanisms related to the
cluster environment, and instead they are most likely to become passive
 {\em before} encountering the cluster environment through galaxy mergers,
 which are most frequent in galaxy groups or cluster infall regions.

The environmental trends of fainter galaxies
\mbox{(M$_{r}\!\ga\!{\rm M}^{*}+1$} mag) have generally been examined using
galaxy colors as a measure of their star-formation history. Whereas
the color of massive galaxies becomes steadily redder with increasing
density, a sharp break in the mean color of faint galaxies is observed at
a critical density corresponding to $\sim\!{\rm R}_{vir}$
\citep{gray,tanaka}. In addition,
the relative fraction of red and blue galaxies and the shape of the
luminosity function of red galaxies over \mbox{${\rm
    M}^{*}\!+1\!\la\!{\rm M}_{r}\!\la\!{\rm M}^{*}\!+6$ mag} change
dramatically with density {\em 
  inside} the virial radius \citep{mercurio}. These results
imply that the evolution of dwarf galaxies is primarily driven by
mechanisms directly related to the structure to which the galaxy is bound, such
as suffocation (whereby galaxies lose their halo gas reservoir to the
host halo), galaxy harassment and/or ram-pressure stripping.

These differences in the environmental trends and evolution of high-
and low-mass galaxies are most likely related to the observed strong
bimodality in the properties of galaxies around a stellar mass of
\mbox{$\sim\!3\times10^{10}{\rm M}_{\sun}$} \mbox{($\sim\!{\rm
    M}^{*}+1$\,mag)}, with more massive galaxies predominately passive
red spheroids, and less massive galaxies tending to be blue
star-forming disks \citep[][hereafter K03a,K03b]{k03a,k03b}. This
implies fundamental differences in the formation and evolution of
giant and dwarf galaxies, and it has been proposed
\citep[e.g.][]{dekel} that this is related to the way that gas from
the halo cools and flows onto the galaxy, which affects its ability to
maintain star-formation over many Gyr. 

We examine the origins of this bimodality by comparing the
environmental dependences of star-formation in giant and dwarf galaxies.
The SDSS Data Release 4
\citep[DR4;][]{sdssdr4} is used to construct a sample of galaxies over
a \mbox{$26\times26\,{\rm Mpc}^{2}$} region containing the
supercluster centered on the rich cluster A2199 at \mbox{$z=0.0309$}
\citep{rines} that is \mbox{$\ga\!9$0\%} complete to
\mbox{M$^*+3.3$}, and for which we estimate the mean stellar
ages and current star-formation rates (SFRs) of galaxies through their
spectral indices. We adopt a cosmology with
\mbox{$\Omega_{M}=0.27$}, \mbox{$\Omega_{\Lambda}=0.73$} and
\mbox{H$_{0}=73$\,km\,s$^{-1}$\,Mpc$^{-1}$} \citep{spergel}. 

\section{The Data}
\label{data}

\begin{figure}
\plotone{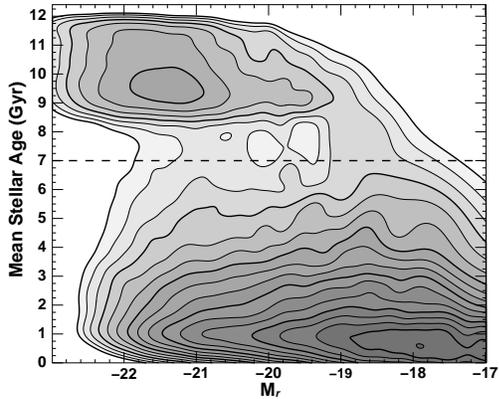}
\caption{Bivariate volume-weighted (1/V$_{{\rm max}}$) number density distribution of all
  $0.005\!<\!z\!<\!0.10$ galaxies as a function of mean stellar age and M$_r$. Contours
  are logarithmically spaced, with the thick contours separated by a
  factor 2 change in number density. The dashed-line indicates the
  boundary defined to separate old and young galaxy populations.}
\label{d4000_Mr}
\end{figure}

The data consists of those galaxies from the SDSS DR4
spectroscopic catalog in the vicinity of the \mbox{$z=0.03$} supercluster,
defined here as the region  
\mbox{$15^{h}56^{m}\!<\!{\rm RA}\!<\!17^{h}00^{m}$},
 \mbox{+34$\degr\!<\!{\rm Dec}\!<\!+46\degr$}, and
 within \mbox{2\,000\,km\,s$^{-1}$} of the redshift of A2199. 
The SDSS spectroscopic magnitude limit of \mbox{$r=17.77$} mag means that at
\mbox{$z=0.0309$} the data covers galaxies with absolute magnitudes
\mbox{${\rm M}_{r}\!<\!-17.8$} mag
\citep[k-corrected to \mbox{$z=0$} using  {\sc kcorrect} v3.2;][]{kcorrect}, or \mbox{${\rm M}_{r}\!<\!{\rm M}^{*}+3.3$} mag
\citep[\mbox{M$_{r}^{*}=-21.15$} mag at \mbox{$z=0.0309$;}][]{blanton03}.
We define two galaxy subsamples from this dataset: a {\em giant} 
subsample of the 546 galaxies with \mbox{M$_{r}\!<\!-20$} mag, which can be
directly compared with previous studies; and a
{\em dwarf} subsample of 777 galaxies with \mbox{$-19\!<\!{\rm
    M}_{r}\!<\!-17.8$} mag. 

We use the stellar indices of the Garching SDSS DR4 catalogs (K03a), in
which the emission-line fluxes are corrected for stellar absorption
using a continuum-fitting code based on the \citet[][hereafter
  BC03]{bruzual} population synthesis models \citep{tremonti}.
Each index/color is corrected for internal dust extinction by adopting
the A$_z$ values of K03a and the $\lambda^{-0.7}$ attenuation law of \citet{charlot}. 
 The $r$-band luminosity-weighted mean
stellar age (hereafter described as the 'age') of each galaxy
is estimated by simultaneously fitting the 
d4000 \citep[as defined in][]{balogh99}, H$\delta_A$, H$\beta$ and
[MgFe]$^\prime$ indices and $(g-r)$, $(r-i)$ colors to BC03
model stellar
populations having exponentially-decaying SFRs with time-scales
between 1 and 10\,Gyr, a Salpeter IMF, and metallicities
0.4--2.5\,$Z_{\sun}$. 
Independently, we also consider the H$\alpha$ emission which can be
directly related to the current SFR \citep[e.g.][]{kennicutt}.

\begin{figure}
\plotone{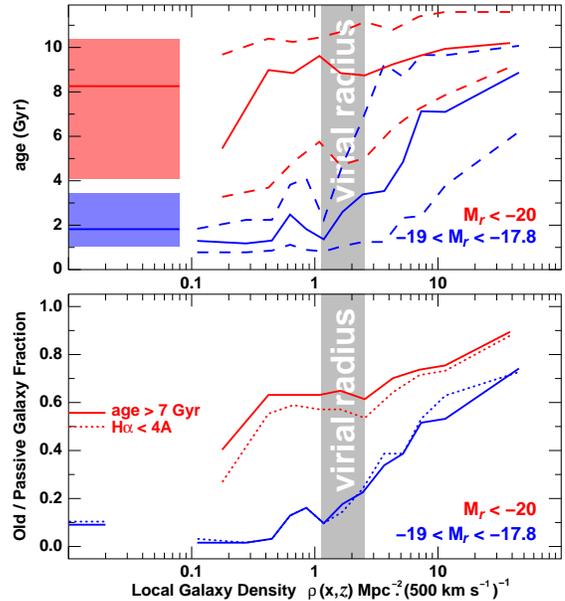}
\caption{{\em Top panel:} The distribution of mean stellar age as a function of
  local galaxy density, \mbox{$\rho({\mathbf x},z)$} for giant (red) and dwarf (blue) galaxies. The
  solid and dashed lines show respectively the median and
  interquartile values of the age distribution in bins of 50
  galaxies. The color-shaded regions on the left indicate the
  corresponding values for the whole SDSS DR4 dataset. 
{\em Bottom
  panel:} The fraction of giant (red) and dwarf (blue) galaxies
with high mean stellar ages \mbox($t\!>\!7$\,Gyr; solid) or little
  ongoing star-formation \mbox{EW$({\rm
  H}\alpha)\!<\!4\,$\AA; dotted)}  as a function of environment. The corresponding fractions for
  the whole SDSS DR4 dataset are shown by the left axis.
}
\label{d4000_rho}
\end{figure}

The local environment is quantified by considering all 1940 galaxies
spectroscopically confirmed as lying within the A2199 region defined above,
including a further 142 galaxies \citep[mostly from][]{rines} not contained within SDSS DR4.
The local galaxy number density, \mbox{$\rho({\mathbf x}, z)$}, is determined for each galaxy
using a variant of the adaptive kernel estimator \citep{pisani96},
whereby each galaxy is represented by a Gaussian kernel in
redshift-space, \mbox{$K({\mathbf x}, z)$}, of width
500\,km\,s$^{-1}$ in the radial direction, and whose
transverse width is set to the distance of its $5^{th}$ nearest neighbor
within \mbox{500\,km\,s$^{-1}$}, a value and approach motivated 
to define the environment of a galaxy on the scale of its host halo
\citep[e.g. K04;][]{yang}.

Galaxy clusters and groups in the region are identified as local maxima in
$\rho({\mathbf x}, z)$, and can be reliably characterized in terms of
their member galaxies,
velocity dispersions and virial radii 
\citep[as in][]{girardi} for groups as poor as \mbox{$\sim\!20$0\,km\,s$^{-1}$}
 using the biweight algorithm \citep{beers}.

\begin{figure*}
\plotone{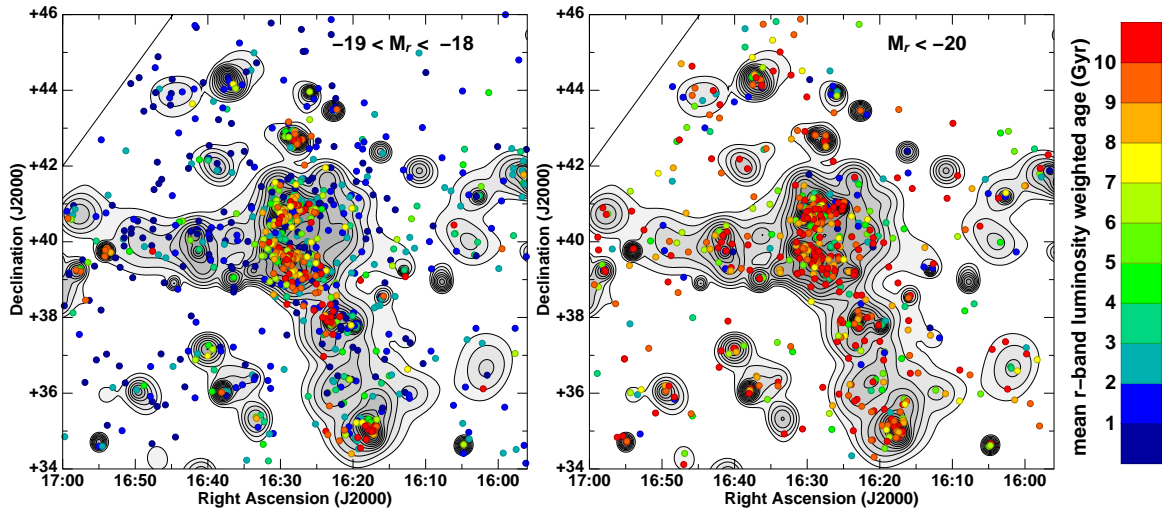}
\caption{The mean $r$-band luminosity weighted stellar age (solid colored circles) as a function of spatial position in the A2199 supercluster environment, for dwarf
  (\mbox{$-19\!<\!{\rm M}_r\!<\!-17.8$ mag;} left) and giant
  (\mbox{${\rm M}_r\!<\!-20$ mag;}
  right) galaxies. The black contours represent the local
  luminosity-weighted surface density of galaxies with redshifts within \mbox{2\,000\,km\,s$^{-1}$}
  of A2199. While old giant galaxies
  are ubiquitous, old dwarf galaxies are found only in the
  densest regions.
}
\label{d4000spatial}
\end{figure*}

\section{Results}
\label{results}
Figure~\ref{d4000_Mr} shows the relation between mean stellar age and M$_r$ for
all \mbox{$0.005<z<0.10$} galaxies in the Garching SDSS DR4 sample. The
distribution is clearly bimodal, with a population of bright
\mbox{($\sim\!L^{*}$)} galaxies \mbox{$\approx\!10$\,Gyr} old, and a population
of fainter galaxies dominated by young \mbox{($\la\!3$\,Gyr)}
stars. We henceforth describe these two populations as old
($t\!>\!7$\,Gyr) and young ($t\!<\!7$\,Gyr) galaxies, and remark
 here the apparent lack of faint (M$_r\!\ga\!-19$ mag) old galaxies.
We stress that ``old'' galaxies may well have been assembled at much
later epochs than when the bulk of their constituent stars formed,
while ``young'' galaxies may also contain 
significant old stellar populations. 
Additionally, a number of systematic effects could affect our age
estimates. These include the choice of model star-formation histories and stellar populations
\citep[e.g.][]{maraston,tantalo}, aperture bias effects, and the impact of
[$\alpha$/Fe] abundance ratio variations, particularly on old metal-rich
populations \citep{thomas04}.
However, we
stress that the same bimodality is
apparent in each of the age-sensitive indices (see e.g. K03b), while
the individual classification of galaxies as young or old, and the
resultant trends with environment are both found to be 
robust for any reasonable choice of model or indices used. 

The dependence on local galaxy density of
the distribution of stellar ages for giant (shown in red) and dwarf
(blue) galaxies is shown in the top panel of Fig.~\ref{d4000_rho}. The solid and dashed lines show respectively the
median and interquartile values of the age distribution. Both giant and dwarf subsamples show a steady increase of age with
local density from field values (color-shaded regions) to the high-density regions where both giant and dwarf
populations are predominately old. In all environments giant galaxies
are on average 1--7\,Gyr older than dwarfs. Moreover 
as density decreases, the age distribution of giant galaxies {\em extends}
to include ever
lower ages, while that of dwarf galaxies gets younger but also {\em
  narrows}, so that at the lowest
densities only young ($\la\!2$\,Gyr) galaxies are found.

The lower panel of Fig.~\ref{d4000_rho} shows the fraction of giant
and dwarf galaxies with \mbox{$t\!>\!7$\,Gyr} as a function of
environment (solid curves). In the
highest-density regions \mbox{$\ga8$0\%} of both giant and dwarf
populations are old. Whereas the fraction of giant
galaxies with \mbox{$t\!>\!7$\,Gyr} declines gradually with decreasing density to the global field
value of $\sim5$0\%, that of dwarf galaxies drops rapidly to \mbox{$\la\!2$0\%}
at densities typical at the cluster virial radius (shown as the shaded stripes in 
Fig.~\ref{d4000_rho}), and continues to
decrease, tending to {\em zero} for the lowest density bins. 
Identical
trends with density are independently obtained (dotted curves) when
passive galaxies are identified from their lack of H$\alpha$
emission \citep[\mbox{EW$[{\rm H}\alpha]\!<\!4$\,\AA};][]{balogh04a}.
\citet{balogh04b} obtain similar trends using red $(u-r)$ colors to
identify passive galaxies, except that a small
fraction \mbox{($\sim\!1$0\%)} of red low-mass galaxies remains even in low-density regions, which could represent
contamination from dusty star-forming galaxies \citep[e.g.][]{wolf}.

To relate the differences in the trends directly to the effect of the 
supercluster, Fig.~\ref{d4000spatial} shows the mean stellar age as a
function of spatial distribution, for the dwarf
(\mbox{$-19\!<\!{\rm M}_{r}\!<\!-17.8$} mag; left panel) and giant
(\mbox{M$_r\!<\!-20$} mag; right panel) galaxy subsamples. Each supercluster galaxy is indicated
by a solid circle whose color represents its age from 
red \mbox{($t\!\ga\!9$\,Gyr)} to blue
\mbox{($t\!\la\!3$\,Gyr)}
, while the black isodensity contours represent the local luminosity-weighed galaxy density.

The giant galaxy population in the centers of clusters or groups are
predominantly old \mbox{($\ga\!9$\,Gyr)}, whereas in regions outside
of any structure, there is a complete range of galaxy ages with an
equal interspersed mixture of both young \mbox{($\la\!3$\,Gyr)} and
old galaxy populations, indicating that their evolution is
driven primarily by their merger history rather than by direct interactions with
their large-scale environment.

In contrast, the mean stellar ages of dwarf galaxies are strongly
correlated with environment: while the cores of the richest
\mbox{($\sigma\!\ga\!50$0\,km\,s$^{-1}$)} clusters are still dominated by
 old \mbox{($\ga\!7$\,Gyr)} galaxies, elsewhere almost all
 \mbox{($\ga\!9$5\%)} of the dwarf galaxies are dominated by young
 \mbox{($\la\!3$\,Gyr)} stars, and of the few remaining old galaxies
      {\em all} are found in
 either poor groups or \mbox{$\la\!20$0\,kpc} from an old giant \mbox{($\ga\!L^{*}$)} galaxy. 

\begin{figure*}
\plotone{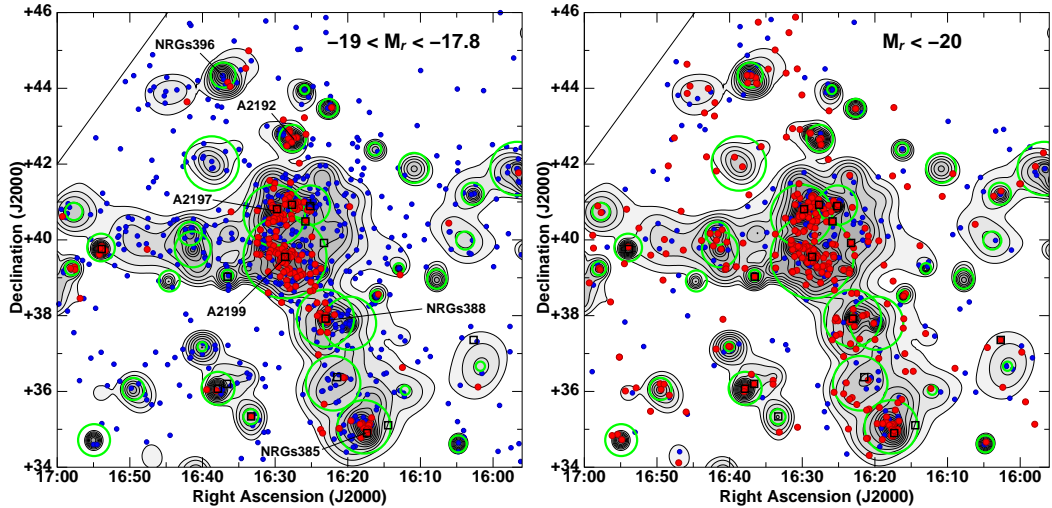}
\caption{The distribution of galaxies with (solid blue circles) and without
  (\mbox{EW(H$\alpha)\le4$\,\AA;} solid red circles) H$\alpha$ emission in the A2199
  supercluster environment, for dwarf
  (\mbox{$-19\!<\!{\rm M}_r\!<\!-18$ mag;} left) and giant
  (\mbox{${\rm M}_r\!<\!-20$ mag;}
  right) galaxies. Black squares indicate \mbox{$r\!<\!13.3$ mag}
  \mbox{($\ga\!3\,L^{*}$)} galaxies. Isodensity contours are shown as in
  Fig.~\ref{d4000spatial}. The green circles indicate the
  virial radii of the galaxy groups/clusters associated with the A2199
  supercluster. }
\label{halpha}
\end{figure*}

As an independent observation of these effects,  Fig.~\ref{halpha}
shows the the spatial distributions of dwarf (left panel) and giant (right) galaxies identified as passive by their
lack of significant H$\alpha$ emission (\mbox{EW$[{\rm H}\alpha]\leq4$\,\AA;} 
solid red circles).

As in Fig.~\ref{d4000spatial}, the spatial distributions of dwarf and
giant galaxies without H$\alpha$ emission are completely different:
whereas the passive giant galaxies are found {\em throughout} the
region covered, passive dwarf galaxies are very strongly concentrated
towards the cluster environments, with \mbox{$\ga\!9$0\%} located
within the virial radius of one of the clusters or groups (shown by green circles). 
In addition, as before, {\em none} of the passive dwarf galaxies are
found to be isolated: those not in one of the clusters or groups are associated with an
$L^{*}$ galaxy, and often these neighbors are more luminous yet, having
\mbox{$r<\!13.3$ mag} or \mbox{$\ga\!3L^{*}$} (shown as open black squares).

\section{Discussion}
\label{discussion}
Using two independent indicators of the past and present
star-formation in galaxies, we present a clear
demonstration of the quite different relationships between
star-formation and environment for giant \mbox{($\sim\!L^*$)} and dwarf \mbox{($\la\!0.1L^*$)} galaxies.
As found in previous studies, giant galaxies \mbox{(M$_{r}\!<\!-20$ mag)}
show a gradual transition from \mbox{$\ga\!8$0\%} being classed as old
\mbox{($t\!>\!7$\,Gyr)} or passive \mbox{(EW$[{\rm H}\alpha]\le4$\,\AA)} in the cluster cores, to field regions
beyond \mbox{3--4\,$R_{vir}$} where still \mbox{$\sim\!4$0\%} are classed as passive.
These results can be understood in the context of massive galaxies
becoming passive through galaxy-galaxy interactions: the finding of both passive and
star-forming galaxies in all environments reflects its stochastic
nature and its independence from large-scale structure. The
gradual overall trend reflects the increasing probability with density
for a galaxy to have undergone a major merger in its lifetime.

In contrast the dwarf galaxy population (\mbox{$-19\!<\!{\rm M}_{r}\!<\!-17.8$} mag) shows a sharp
transition at \mbox{$\sim\!R_{\rm vir}$}, from being predominantly
old and passive within the cluster, to outside where virtually all galaxies
are young and forming stars, and
passive galaxies are {\em only} found as satellites to more massive
galaxies.
These findings are supported by the observed dependencies of galaxy
clustering on luminosity and color, in which the galaxies associated
with the most overdense regions on $\la\!1$\,Mpc scales are found to be both the bright \mbox{(${\rm
    M}_{r}\!<\!-22$ mag)} and faint \mbox{(${\rm
 M}_{r}\!>\!-19$ mag)} red galaxies \citep{hogg,zehavi}.  
These dependencies were
successfully reproduced by the smoothed particle hydrodynamics cosmological
simulations of \citet{berlind}, who identify the passive low-mass
galaxies as satellites within more massive halos.

The observation that no passive dwarf galaxies are found in isolated
(with respect to \mbox{$\ga\!L^{*}$} galaxies) low-density regions
implies that galaxy merging cannot be effective in completely terminating
star-formation in low-mass galaxies, particularly as in this study these
regions represent the infall regions of the supercluster where low-velocity
encounters should be most frequent. 
Low-mass galaxies become passive only when they become
satellites within a more massive DM halo and their halo gas reservoir
is lost to that of the host, ``suffocating'' the galaxy \citep{larson,bekki}, or
through directly cluster-related mechanisms such as galaxy harassment
and/or ram-pressure stripping.

These differences can be understood in the context of the hot and cold
gas infall \citep{dekel,keres} or AGN feedback \citep[e.g.][]{croton,hopkins}
models of galaxy evolution.
When gas-rich galaxies merge, tidal forces 
 trigger a starburst and fuel the rapid growth of the central black
 hole, until outflows from the AGN drive out the remaining cold gas
 from the galaxy, rapidly terminating the starburst. In massive
 galaxies, the gas in the halo is also heated by stable virial
 shocks, and is prevented from cooling by feedback from quiescent
 accretion of the hot gas onto the black hole,
 effectively shutting down star-formation \citep{croton}. Because black hole growth
 is strongly dependent on galaxy mass, AGN feedback in low-mass
 galaxies is much less efficient at expelling cold gas or affecting
 star-formation \citep{springel}. 
In addition, low-mass galaxies self-regulate their star-formation
through supernova feedback, preventing starbursts that exhaust the cold
gas, which in turn is constantly replenished
by cold streams, so that their ability to maintain star-formation over
many Gyr is much less affected by their merger history.

\acknowledgements{The authors thank the anonymous referee for useful
  comments which helped us to significantly improve the article.
CPH acknowledges the financial supports provided
  through the European Community's Human Potential Program, under
  contract HPRN-CT-2002-0031 SISCO. 
This work is partially supported by the Italian Ministry of Education,
  University and Research (MIUR) grant COFIN200420323: {\it The
  Evolution of Stellar Systems: a Fundamental Step Towards the
  Scientific Exploitation of VST}.
Funding for the creation and
  distribution of the SDSS Archive has been provided by the Alfred
  P. Sloan Foundation, the Participating Institutions, the National
  Aeronautics and Space Administration, the National Science
  Foundation, the U.S. Department of Energy, the Japanese
  Monbukagakusho, and the Max Planck Society. 
}

\label{lastpage}
\end{document}